\documentclass[aps,final,notitlepage,oneside,twocolumn,nobibnotes,nofootinbib,
superscriptaddress,noshowpacs,centertags]{revtex4-1}

\usepackage[utf8]{inputenc}
\usepackage[english]{babel}
\usepackage{graphicx}
\usepackage{latexsym}
\usepackage{amssymb}
\usepackage{amsmath}
\usepackage{float}

\begin{document}

\title{Triggering star formation by free-floating planets}
\author{Yu. N. Eroshenko}\thanks{e-mail: eroshenko@inr.ac.ru}
\affiliation{Institute for Nuclear Research of the Russian Academy of Sciences, 60th October Anniversary Prospect 7a, Moscow, 117312, Russia}

\date{\today}

\begin{abstract}
Recently, a population of free-floating planets has been discovered in the Galaxy. This paper examines the influence of the planets trapped within protostellar hydrogen clouds on the processes of gravitational instability in the gas. A sufficiently massive planet can become an alternative dynamic center for the contracting cloud. This occurs due to rapid mass growth from accretion onto the planet within the cloud. Owing to residual velocity and density gradients in the cloud, the planet that triggers star formation is likely to end up in a distant orbit around the newborn star. Some second- and subsequent-generation stars in the Galaxy may have formed via this mechanism.
\end{abstract}


\maketitle

\section{Introduction}

According to modern concepts, most stars are formed in cold hydrogen clouds. After the fragmentation of the cloud stops, the stage of energy emission and monotonous compression begins \cite{HayNak65}. At this stage, the contracting cloud must choose a dynamic center, in the direction of which compression occurs, and where a star is eventually formed. The dynamic center in a normal situation, obviously, should roughly correspond to the geometric center of the cloud, its center of mass. In this paper, we consider an atypical situation when there is a free-flying planet in the cloud, which, even though it is not in the center of the cloud and has a mass several orders of magnitude less than the cloud mass, may at some time accept the role of a dynamic center. The region of star formation can include hundreds and thousands of stars at once. But over time, these stellar associations should disperse \cite{Tut78}. Planets can fly out of protoplanetary disks \cite{TutDreDre20} or at later stages of evolution \cite{AndPop21}. These free-flying planets \cite{Mroetal17,PerKou12,MusRayDav16,ParLicQua17,DafParQua22,Donetal26} could have remained within the star formation region or been trapped in a collapsing cloud \cite{Ero23}. 

A planet located in a protostellar cloud can accept the role of a dynamic center due to the gas accretion on it. As a result, gas condensation is rapidly increasing, which eventually becomes an alternative center of gravity in the entire cloud. According to our calculations, a planet with a mass of the order of Jupiter's mass $M_{\rm pl}\sim M_{\rm J}$ manages to pull on enough gas during the compression of the entire cloud to become a dynamic center. At the same time, due to the influence of dynamical friction, the planet is effectively slowed down relative to the surrounding gas and always remains within the contracting cloud. More precisely, due to the planet's motion and inhomogeneities, it will end up not exactly in the center, but in a wide orbit around the formed star. 

In the 1950s, \cite{Kra52,Ure56,Hua57} hypothesized that stars were formed as a result of gas accretion to pre-existing solid planets, the origin of which remained unclear (the possibility of dust condensation was discussed). In this paper, we consider another scenario in which planets are responsible for star formation, but these planets are not located in the center of the forming star. On the contrary, it is more likely that the planet only accepts the role of the dynamic center of a contracting gas cloud for a while, but eventually finds itself in orbit around the star outside the planetary disk.

There is an increasing amount of observational data on unusual exoplanetary systems, the formation of which is difficult to explain within the framework of conventional models of protoplanetary disk formation. For example, about a third of the Sun-class stars have planetary disks that are not aligned with the rotation of the star \cite{Bidetal25}, see especially \cite{Moretal25}, \cite{Bay25}. A giant planet $0.17M_{\rm J}$ has been discovered around a very light star with a mass of $0.2M_\odot$\cite{Bryetal25}. Previously, it was believed that the protoplanetary disk around a light star is too small and giant planets cannot form in it. Even in the solar system, it is not clear why angular momentum is mostly contained in the planets and not in the Sun itself. These and other examples may require non-standard models of star formation. A model with a planet as an alternative dynamic center may be useful for explaining some systems with a rare configuration.


\section{Compression of a protostellar cloud}

Let's assume that at some initial moment $t_i$ there was a spherical cloud of neutral hydrogen at rest.  The values at the time $t_i$ are then labeled with the index ``i''. We assume that cloud compression begins from the moment $t_i$. For example, we will choose the initial cloud parameters based on the typical parameters specified in \cite{Sur01}.  The mass of the cloud is $M_g\simeq 1M_\odot$, the number density of the hydrogen is $n_i\simeq10^5$~cm$^{-3}$, the density is $\rho_i=n_im_p$, where $m_p$ is the mass of a proton (neglecting the contribution of helium), the temperature of the gas is $T_i\simeq10$~K. Such a cloud has the initial radius of $R_i=(3M_g/(4\pi\rho_i))^{1/3}\simeq9.5\times10^3$~AU. In \cite{Sur01} was also indicated that the cloud is compressing isothermically. The cloud releases energy and is compressed in free fall mode until it is compressed by about 100 times in radius. After that, opacity sets in and the processes of gas heating, star formation, and protoplanetary disk formation are initiated. 

Let's consider the specified compression stage in the free fall mode, neglecting the density inhomogeneity. The evolution of the cloud radius $R(t)$ obeys the simple equation 
\begin{equation}
\frac{dR}{dt}=-\frac{GM_g}{R^2}.
\end{equation}
This equation has the known solution in a parametric form
\begin{equation}
R=R_i\cos^2\theta,
\label{sol1}
\end{equation}
where the time dependence 
\begin{equation}
t-t_i=\left(\frac{3}{8\pi G\rho_i}\right)^{1/2}\left(\theta+\frac{1}{2}\sin(2\theta)\right).
\label{sol2}
\end{equation}
Since compression started from a state of rest, at the initial moment $dR/dt=0$ (this corresponds to $\theta=0$). We denote by $t_c$ ($\theta=\pi/2$) the moment of formal compression of the cloud to a point according to the solution (\ref{sol1}), (\ref{sol2}). In reality, this moment is not reached, because the opacity of the cloud \cite{HayNak65} sets in, the gas heats up, its pressure becomes important, and the solution (\ref{sol1}), (\ref{sol2}) loses its validity.

We are interested in the very last stage of compression, but before opacity occurs. In this case, we can write $\theta=\pi/2+\varepsilon$ and expand by the small parameter $\varepsilon$. By doing this calculation, we obtain for the radius
\begin{equation}
R(t)=R_i(6\pi G\rho_i)^{1/3}(t_c-t)^{2/3}.
\label{rtlast}
\end{equation}
From this we get that the stage of opacity that occurs at $R\simeq0.01R_i$ corresponds to the time of $t_c-t\simeq69$~years. The density of the cloud varies according to the law
\begin{equation}
\rho(t)=\rho_i\left(\frac{R_i}{R}\right)^3=\frac{1}{6\pi G(t_c-t)^2}.
\label{rhosol}
\end{equation}


\section{Perturbation theory}

Using the theory of small perturbation growth, we will demonstrate now that a planet with a mass of $M_{\rm pl}\sim M_{\rm J}$ creates a density perturbation in the proptostellar cloud $\delta=\delta\rho/\bar\rho$, which increases to a magnitude of $\delta\sim1$ to the onset of cloud opacity. I.e., such a planet can take the role of a new dynamic center of the contracting cloud.

The evolution of small perturbations in a compressing medium was investigated in \cite{Hun62} (for a modern approach, see \cite{Tocetal18}). Using rather cumbersome calculations, it was shown that the law of perturbation growth has the form (in our notations) $\delta\propto1/(t_c-t)$. We can reproduce this result from the simple arguments. Indeed, the evolution of a homogeneous cloud is mathematically equivalent to the evolution of a homogeneous cosmological model with time reversal, when the Hubble parameter $H=\dot R/R<0$. Let us write down the well-known equation of the evolution of perturbations in cosmology
\begin{equation}
\frac{\partial^2\delta}{\partial t^2}+2H\frac{\partial\delta}{\partial t}=4\pi G\bar\rho\delta.
\label{deleq}
\end{equation}
Then for $H=-2/[3(t_c-t)]$ we immediately get the specified solution $\delta\propto1/(t_c-t)$.

At the initial moment, consider the sphere   with a radius of $\sim R_i/2$ surrounding the planet and located on one side of the initial dynamic center. The mass of the gas within this sphere is $\sim M_g/8$. The planet creates in this gas initial perturbation $\delta_i\sim 8M_{\rm pl}/M_g$. For the mass parameters of the Sun and Jupiter $\delta_i\sim0.008$.  The asymptotics (\ref{rtlast}) and (\ref{rhosol}) are applicable when $R\ll R_i$. For a conservative estimate, let's take the initial value at $R\sim R_i/5$. Compression to opacity corresponds to $R\sim R_i/100$. Then we get that during the time between these two events, the perturbation $\delta$ increases according to the law $\delta\propto1/(t_c-t)$ up to a nonlinear value $\delta\sim0.8$. If we take into account the initial stage of the compression, the growth will be even greater. This means that even a conservative estimate suggests that the planet may become a dynamic center, or an alternative center of gas condensation, in which a second star of the binary system will born. And for a more massive planet, the effect is even stronger. In the next section, we will further justify this conclusion from the accretion formalism.


\section{Accretion of gas onto the planet}

\begin{figure}[h]
\begin{center}
\includegraphics[width=0.49\textwidth]{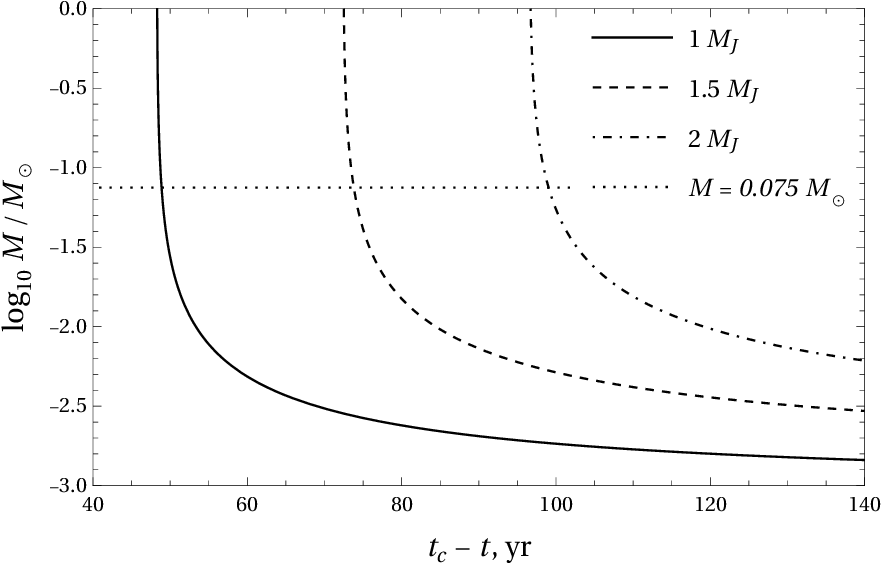}
\end{center}
\caption{The mass of the accreted gas depending on the time before the collapse of the cloud for three exoplanet masses 1, 1.5 and 2 Jupiter masses. The point curve shows the minimum mass of a star.} 
\label{gr1}
\end{figure}

Let us now consider the accretion of gas onto a planet falling towards the center along with a contracting hydrogen cloud. We assume that the planet has been within the cloud for a long time, so its velocity due to dynamic friction is almost zero. The dynamic friction is considered in the next Section. Due to the gas density increase, the accretion rate increases sharply  at the last stage, and a mass comparable to the mass of the entire protostar can accumulate around the planet. Mass growth during Bondi accretion \cite{Bon52}  occurs as
\begin{equation}
\frac{dM}{dt}=\frac{\pi\rho(t)G^2M^2}{c_s^3},
\label{acceq}
\end{equation}
where the sound speed of molecular hydrogen is $c_s=[7k_{\rm B}T_i/(5(2m_p))]^{1/2}$, $k_{\rm B}$ is the Boltzmann constant. Substituting (\ref{rhosol}) and solving this equation, we get
\begin{equation}
M(t)=\frac{M_i}{1-\left(GM_i/(6c_s^3)\right)(t_c-t)^{-1}},
\label{mgassol}
\end{equation}
where $M_i=M_{\rm pl}$. The function (\ref{mgassol}) is shown at Fig.~\ref{gr1} for the three masses $M_{\rm pl}$ of the exoplanet. The accreted gas accumulates around the planet in the form of a dense condensation. The greater its mass, the stronger the gravitational field and the higher the accretion rate. For $M_{\rm pl}\geq1.5M_{\rm J}$, where $M_{\rm J}$ is the mass of Jupiter, the accreted mass becomes comparable to the mass of the entire cloud before opacity occurs. This means that at the stage of transparency (isothermal free fall), the exoplanet accepts the role of a dynamic center, and the star is formed not in the dynamic center of the initial cloud, but near the gas condensation that the planet created. For smaller masses of exoplanets, this mechanism does not work, and they do not become dynamic centers, but simply enter the emerging planetary system as one of the planets. But, perhaps, with an abnormal chemical composition.

Even if the condensation around the planet does not reach the mass of the entire cloud, it can reach the minimum mass of $\sim0.075M_\odot$, at which thermonuclear reactions are ignited and a star is formed. In this case, it is possible to form a binary system in which the main star was formed by the compression of the main cloud, and the second star was formed from the condensation of gas around the planet.

One can also find the Bondi radius $r_{\rm B}=GM(t)/c_s^2$. It can be seen that even before the planet became a dynamic center, the Bondi radius exceeded the current radius of the cloud. Under this condition, accretion occurs in supersonic mode, gas flows onto the planet in free fall regime, and a large mass rapidly builds up around the planet.

In this section, accretion in a contracting molecular cloud is considered. Note that the typical growth time of the mass condensation around the planet in a stationary cloud is much longer. If we put $\rho=const$ in (\ref{acceq}), then instead of (\ref{mgassol}) we get the solution $M(t)=M_i/(1-t/\tau)$, where the characteristic time is
\begin{eqnarray}
\tau &=& \frac{c_s^3}{\pi \rho G^2M_{\rm pl}} 
\\
&\simeq&10^8\left(\frac{\rho}{1.7\times10^{-19}\mbox{g~sm$^{-3}$}}\right)^{-1}\left(\frac{M_{\rm pl}}{M_{\rm J}}\right)^{-1}\mbox{~yrs}.
\nonumber
\end{eqnarray}
This is the time when instability is initiated by the planet in a stationary cloud that has not yet begun to contract.


\section{Dynamical friction}
\label{dfsec}

When switching to the opacity mode of the cloud, the gas begins to slow down by pressure. However, for a planet, as can be shown, the drag of the gas is negligible. This means that the planet could have flown out of the contracting cloud of gas. However, the dynamical friction effect predicted by Chandrasekhar will not allow a planet with a sufficiently large mass to fly out of the cloud. Dynamic friction consists in the fact that a moving planet creates a density disturbance behind it, and this disturbance attracts the planet with its gravitational field and slows down its movement.  Let's take a closer look at this process.

According to the Chandrasekhar formula, the dynamical friction force for a planet moving at a velocity of $v$ has the form
\begin{equation}
F_{\rm df}=-4\pi \left(\frac{GM_{\rm pl}}{c_s}\right)^2\Lambda(X\sqrt{2})^{-2}\left[{\rm erf}(X)-\frac{2X}{\sqrt{\pi}}e^{-X^2}\right],
\label{df}
\end{equation}
where $X=v/{\sqrt{2}\sigma}$, $\sigma$ is the particle velocity dispersion, and $\Lambda$ is the Coulomb logarithm. This expression assumes that the environment is collisionless. In \cite{Ost99}, Chandrasekhar's approach was generalized to the case of a pressurized gas medium. As a result, the Coulomb logarithm and the expression in square brackets take on a slightly different form. In \cite{EscLarCop04}, numerical simulation was used to verify the predictions of \cite{Ost99}. The result of this test was that the force of dynamical friction in a gas can be described by the same formula (\ref{df}) if one put $\Lambda=1.5$ for $X\sqrt{2}\leq0.8$ and $\Lambda=4.7$ for $X\sqrt{2}>0.8$, as well as $\sigma=c_s$. We use this recipe in our calculation of the motion of the planet relative to the gas cloud. 

First of all, the effective deceleration time (speed reduction by a factor of $e$) has the form (in $\Lambda=const$ approximation)
\begin{equation}
T_{\rm df}=\frac{3c_s^3}{2^{7/2}\pi^{1/2}\rho G^2 M_{\rm pl}\Lambda}.
\label{tdf}
\end{equation}
Numerically
\begin{equation}
T_{\rm df}=3\times10^7\left(\frac{\rho}{1.7\times10^{-19}\mbox{g~sm$^{-3}$}}\right)^{-1}\left(\frac{M_{\rm pl}}{M_{\rm J}}\right)^{-1}\mbox{~yrs}.
\label{tdf2}
\end{equation}
If a planet with the mass of Jupiter was located within this cloud (more precisely, a more extensive cloud that fragmented into clouds with masses of $\sim M_g\sim M_\odot$) for more than $3\times10^7$~years, then it managed to slow down. In this paper, we assume this case and assume that the initial velocity of the planet relative to the cloud is $\ll c_s$. In this case, both the gas of the cloud itself and the planet begin to move towards the initial dynamic center in free fall mode.

\begin{figure}[h]
\begin{center}
\includegraphics[width=0.49\textwidth]{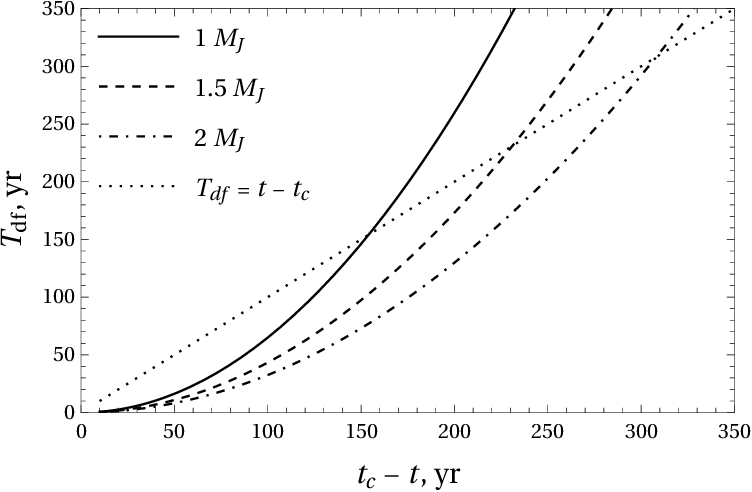}
\end{center}
\caption{Characteristic time of dynamic friction as a function of time before cloud collapse for three exoplanet masses 1, 1.5 and 2 Jupiter masses. The dotted curve corresponds to the equality of these times.} 
\label{gr2}
\end{figure}

As the cloud contracts, its density increases, and the characteristic time of dynamic friction decreases. The time of dynamic friction (\ref{tdf}) as a function of time is shown in Fig.~\ref{gr2}. It can be seen from the graph that many years before the onset of the opacity stage, the characteristic time of dynamic friction becomes less than the remaining time of the cloud evolution. This means that when the cloud passes to the deceleration stage, the planet does not fly out of it, but is effectively slowed down by dynamic friction and remains in the cloud. As a result, it remains within the boundaries of the emerging planetary system. This process can be significantly enhanced by the fact that a local disturbance in the gas density builds up around the planet. This clump of gas is drawn toward the center by the general flow of the gas cloud and at the same time exerts dynamic friction on the planet. As a result, the planet becomes self‑captured by the contracting gas cloud. Nevertheless, the residual velocity most likely leads to the fact that the planet does not fall exactly into the old dynamic center of the cloud and does not become the center of the forming star, but moves into orbit around the star. 

It is clear that this is a non-standard way of star and planetary system formation, therefore, in this case, the parameters of the orbit of a giant planet may also be very atypical. However, our approximate approach does not allow us to obtain numerical values of these parameters. For more accurate conclusions, numerical gas-dynamic simulation of the processes with high spatial resolution is necessary.


\section{Conclusion}

The process of star formation from molecular gas clouds is quite complex and includes the cooling, compression and fragmentation \cite{HayNak65,Sur01}. These processes can be taken into account consistently enough only in three-dimensional hydrodynamic simulations. One of the elements is the development of instability in the gas cloud -- an increase of inhomogeneities. The natural cause of the inhomogeneities is the initial asymmetry of the cloud shape and density gradients formed by shock waves and heating by radiation from neighboring stars. However, these inhomogeneities occur only on the largest scales, initially on the scale of the entire cloud. In the early stages, the Jeans length exceeds the size of the cloud, so all the inhomogeneities on a small scale are quickly smoothed out, they scatter in the form of sound waves and dissipate. As a result, when reducing the Jeans length only small density gradients are involved. Most stars probably formed according to this scenario. First of all, the first generation of stars. 

However, it cannot be ruled out that other types of inhomogeneities could be present in the gas clouds, such as small-scale inhomogeneities created by free-flying planets or primordial black holes. In this regard, the question arises, can these planets initiate instability? Could this be a subdominant mechanism for the formation of next-generation stars? The theory of star formation around solid planetary bodies proposed many decades ago (bottom-up mechanism) \cite{Kra52,Ure56,Hua57} obviously turned out to be untenable. It was found that, on the contrary, planets form in protoplanetary disks around stars formed during the fragmentation and collapse of gas clouds. However, the recently established presence of a population of free-flying planets \cite{Mroetal17,PerKou12,MusRayDav16,ParLicQua17,DafParQua22}, lost during the evolution of stars and planetary systems, suggests that the bottom-up mechanism can indeed be implemented in some small number of cases for planetary systems of the second and subsequent generations. 

In this article, we show that if a free-flying giant planet falls into a cold molecular cloud, it can be slowed down by dynamic friction. Even before the general cloud compression begins, the Bondi accretion of gas onto the planet can lead to the growth of a gas halo around it, and the planet will become a dynamic center for star formation. An even more efficient process of turning a giant planet into a dynamic center is possible in the process of isothermal compression of the cloud. In this case, due to the increasing density of the gas, Bondi accretion accelerates significantly, and the giant planet accumulates a massive halo around itself and becomes a dynamic center even before the end of the general isothermal compression and the onset of opacity of the gas. Although the planet accepts the role of a dynamic center, due to the presence of angular momentum in the cloud, the planet may end up in orbit outside the formed star. This leads to peculiarities in the distribution of angular momentum between the star and the planets. The details of these processes can be traced in high-resolution numerical modeling, which is beyond the scope of this article.

It would be interesting if the 9th planet \cite{Batetal19} could be a trigger for the formation of the solar system. Therefore, according to this model, the Planet~9 has not been expelled from the inner solar system, but it came to us from the protostellar cloud \cite{MusRayDav16,ParLicQua17}. The planet, which was rejected by the builders of previous planetary systems, has become the cornerstone for our solar system formation process. But our estimates show that this scenario is impossible. Only planets with a mass of one and a half times that of Jupiter and more can accept the role of the dynamic center of a contracting protostellar cloud. The primordial black hole could also act as a planet taking on the role of a gravitational center. However, in this case, a problem arises with the fact that primordial black holes are usually distributed in the Galactic halo and have high velocities of $\sim200$km~s$^{-1}$, which prevents their capture into the protostellar cloud \cite{Ero23}.

\bibliography{erobib}

\end{document}